# Beating signals in CdSe quantum dots measured by low-temperature 2D spectroscopy

Zhengjun Wang,[1] Albin Hedse,[1] Edoardo Amarotti,[1] Nils Lenngren,[1,2] Karel Žídek,[1,3] Kaibo Zheng,[1,4] Donatas Zigmantas,[1] Tõnu Pullerits*[1]

**AFFILIATIONS**

[1] Division of Chemical Physics and NanoLund, Lund University, P.O. Box 124, 22100 Lund, Sweden.

[2] ELI Beamlines, Institute of Physics of the Czech Academy of Sciences, Za Radnicí 835, 252 41 Dolní Břežany, Czech Republic.

[3] Regional Center for Special Optics and Optoelectronic Systems (TOPTEC), Institute of Plasma Physics of the Czech Academy of Sciences, 182 00 Prague 8, Czech Republic.

[4] Department of Chemistry, Technical University of Denmark, DK-2800 Kongens Lyngby, Denmark.

*Corresponding author: Tonu.Pullerits@chemphys.lu.se

## ABSTRACT

Advances in ultrafast spectroscopy can provide access to dynamics involving nontrivial quantum correlations and their evolutions. In coherent 2D spectroscopy, the oscillatory time dependence of a signal is a signature of such quantum dynamics. Here we study such beating signals in electronic coherent 2D spectroscopy of CdSe quantum dots (CdSe QDs) at 77 K. The beating signals are analyzed in terms of their positive and negative Fourier components. We conclude that the beatings originate from coherent LO-phonons of CdSe QDs. No evidence for the quantum dot size dependence of the LO-phonon frequency was identified.

## INTRODUCTION

Semiconductor nanocrystals are called quantum dots (QDs) if their size is smaller than the Bohr radius of excitons[1] The availability of simple colloidal synthesis providing well monodispersed QDs[2,3] and the possibility to make stable and highly luminescent core–shell structures[4,5] have made these materials attractive for optoelectronic applications. They are already used to improve the brightness and color gamut of displays[6] and show promise for LED[7] and photovoltaic applications.[8]

QDs have shown potential for future quantum technologies as single photon sources[9] or entangled photon pair sources.[10] Recently they have been discussed as a promising material for applications that rely on the coherent evolution of quantum states.[11,12] Such coherent evolution can be observed as oscillating signals in spectroscopic experiments. Electronic coherent 2D spectroscopy (2DES) has proven to be a particularly efficient technique for revealing such signals.[13–16]

2DES[17–20] is a nonlinear spectroscopic technique that uses three short laser pulses to generate third order polarization in the sample. The first two pulses act as excitation and the Fourier transform over the time delays between them provide the excitation frequency. The signal field emitted by the polarization (the material response) after the third pulse is mixed with a local oscillator field and is spectrally resolved in a spectrometer, thus providing the detection frequency. Such heterodyne detection provides full phase information of the complex-numbered 2D signal as a function of its excitation and detection frequencies. The delay between the first pulse pair (excitation) and the third pulse (detection) is called the population time. If the two excitation pulses leave the system in a coherent superposition of states, beating signals may appear during the population time. The oscillation frequency of such beating is equal to the energy difference of the states that are involved in the superposition.

The states involved in the coherent superposition can be electronic or vibrational. They can also be of mixed electronic-vibrational origin, the so-called vibronic states. Correspondingly, the oscillatory signal reports either electronic, vibrational, or vibronic coherences. How to identify the origin of a certain oscillatory signal in 2D spectroscopy is currently an active discussion in scientific literature.[21–30] It has been shown that the so-called frequency maps obtained via Fourier transform of the whole 2D spectrum over the population time have different patterns. This difference is dependent on whether the beatings originate from electronic or vibrational coherences, in some cases allowing for their origin to be identified. Also, since the 2D spectra are complex and numbered, the positive and negative frequency components of the Fourier transform and the corresponding frequency maps provide complementary information. This gives additional means of distinguishing whether the origin of the beating signal is vibrational or electronic.[23,31]

The size dependence of QD excitonic states can be predicted by effective mass models[32] and were measured in detail for CdSe by Norris and Bawendi.[33] For most QD sizes and combinations of states, the difference in energy between two states increases when the QDs decrease in size, with deviations created by the nonparabolicity of the conduction band and anticrossings of hole levels.[34]

For the phonons, the QD size dependence can be readily explained as the effect of excluding the low k values since the long wavelength phonons do not fit into a QD – an effect called phonon confinement.[35] For the LO phonons in CdSe QDs the frequency shifts around 10 cm$^{-1}$ due to the size change have been reported.[36] LA modes too depend on the size and as expected from the LO and LA phonon dispersion the frequency shifts of the optical and acoustic phonons have opposite sign – while the QD size is reduced the LO phonon frequency is reduced and the LA phonon frequency increased.[37]

In this article, we analyze beating 2D signals of CdSe QDs measured at 77 K. The beating map analyses clearly identify a component of about 220 cm$^{-1}$ as coherent phonons. Since no evident QD size dependence of the LO phonon frequency was identified, such dependence is smaller than the resolution of the current experiment.

**RESULTS AND DISCUSSION**

The sample preparation and the 2DES setup have been described in earlier work.[38,39] For the details, we refer readers to the SI. The 2D spectra were measured at 77 K for the population times up to 20 ps (see Fig. S1 in SI) with suitably chosen step size – denser at early times and with longer steps later on. The scan does not show any significant Auger-effect related faster decay component indicating that no evident charging of the QDs occurs in the experiment even though the sample was not moved during the measurement time. The average number of excitations per QD $<N> = 0.12$.[38] The total, rephasing, and non-rephasing 2D spectra of the CdSe QDs at the population time of 900 fs are shown in Fig. 1. The features of such 2D spectra can be described in terms of quantum pathways, as stimulated emission (SE) and ground-state bleach (GSB), which contribute to the positive signal, and excited-state absorption (ESA), which contributes to the signal with a negative sign.[40–44]

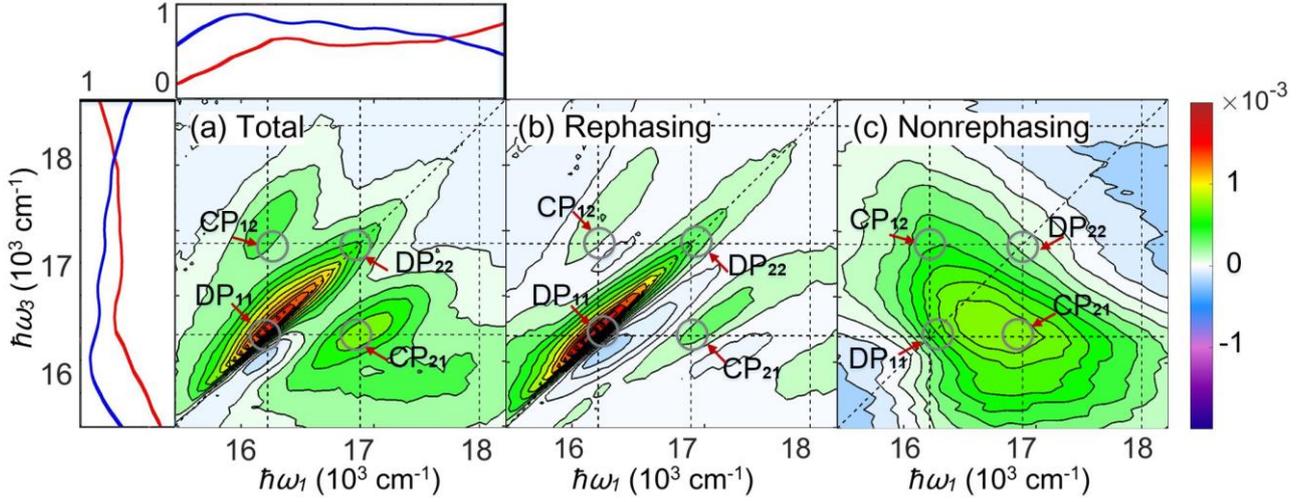

**FIG. 1.** The real parts of the 2D spectra at 900 fs. (a) Total, (b) rephasing, and (c) non-rephasing 2D spectrum. To the left and above the panel (a) the laser intensity (blue) and the CdSe QD absorbance (red) are shown. We have indicated places for several diagonal peaks (DP) and cross peaks (CP).

Based on the analyses of the diagonal of the 2D spectra (for the details, see SI) and the earlier studies of the CdSe QD electronic structure[33] we have marked two diagonal peaks ($DP_{11}$ and $DP_{22}$) and two cross peaks ($CP_{21}$ and $CP_{12}$) in Fig. 1. Since nonrephasing spectra are broadened along the antidiagonal, the four features are not separable in this case. $DP_{11}$ and $DP_{22}$ peaks correspond to the excited states $|e_1\rangle$ at $16200$ cm$^{-1}$ and $|e_2\rangle$ at $16900$ cm$^{-1}$. $|e_1\rangle$ and $|e_2\rangle$ represent the excited states $1S_{3/2}(h) - 1S(e)$ and $2S_{3/2}(h) - 1S(e)$.[33] The $DP_{11}$ peak of the rephasing signal is strongly elongated along the diagonal due to the distribution of the sizes of the CdSe QDs and the corresponding size-dependent inhomogeneous broadening of the involved electronic excited states.

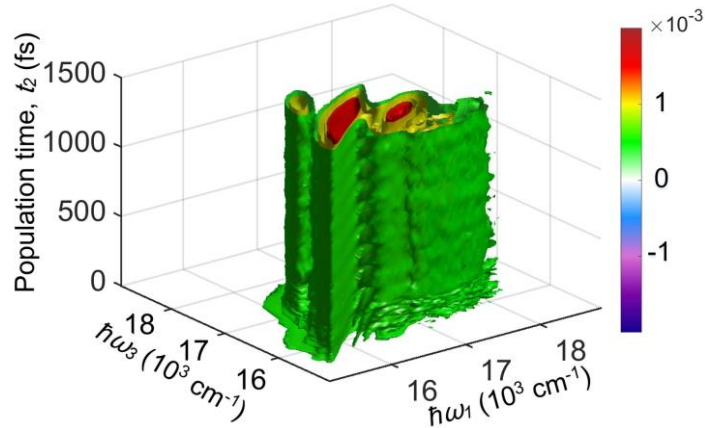

**FIG. 2.** The amplitude of the rephasing 3D Spectra $S(\omega_1, t_2, \omega_3)$ of CdSe QDs

By stacking the 2D spectra as a function of population time we can form a 3D body representing a spectrogram with two frequency and one time dimension, see Fig 2. The body clearly demonstrates an oscillatory time dependence. Such beating signals originate from the coherent excitation of pairs of quantum states. The time evolution of such a superposition of states oscillates with the frequency which corresponds to the energy difference of the two quantum states. The states may be vibrational,[45] electronic[46] or they may have a mixed electronic-vibrational character.[47] The Fourier transform along the population time $t_2$ provides two obvious 3D spectral bodies at around $\pm 200$ cm$^{-1}$ shown in Fig. 3. The bodies are elongated along the diagonal of the ($\omega_1$, $\omega_3$) plane because of the size-related inhomogeneous broadening of the electronic transition energy. Based on the projections of the spectral bodies to the ($\omega_1$, $\omega_2$) and ($\omega_3$, $\omega_2$) planes we can conclude that within the uncertainty of the experiment the beating frequency depends on neither the excitation nor the detection frequency. Since the transition (excitation and detection) frequency can be related to the size of the QDs, in this way we can investigate the QD size dependence of the oscillatory signal frequency.

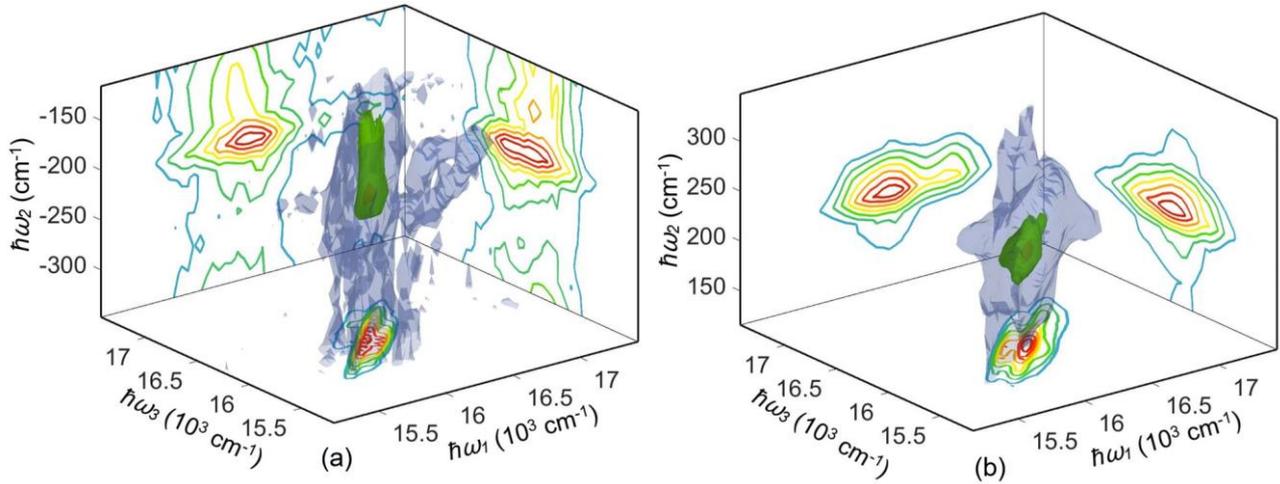

**FIG. 3.** The rephasing 3D spectral bodies of the CdSe QDs after taking a Fourier transform over the population time. The negative (a) and positive (b) $\omega_2$ contain complementary information and are shown separately.

For clarity, we have analyzed the beatings and the corresponding Fourier transform of $DP_{11}$, see Fig. 4.

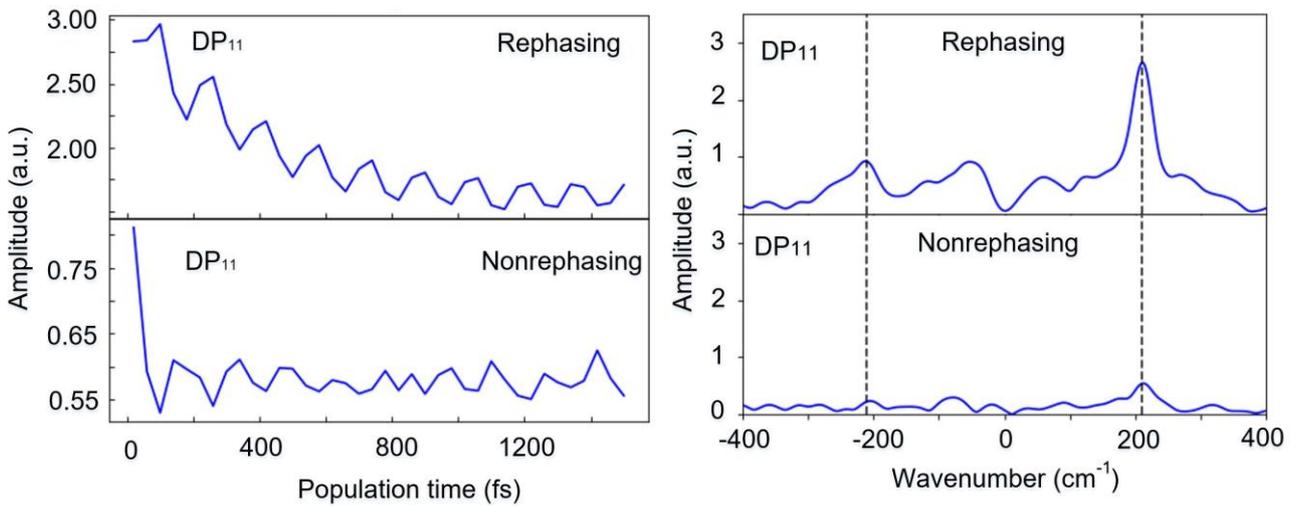

**FIG. 4.** Left: The population time dependence of the rephasing (upper) and nonrephasing (lower) signal at the position of $DP_{11}$. Right: the Fourier transform of the time domain signal, the peaks at ±210 cm$^{-1}$ are marked by a vertical dashed line.

Both positive and negative frequencies show a clear band at 210 cm$^{-1}$. The frequency is consistent with the known LO phonon mode of CdSe QDs[48,49]. In the following, we analyze the positive and negative rephasing 2D frequency maps at 220 cm$^{-1}$ (the closest frequency to 210 cm$^{-1}$ of our 2D maps) to verify the vibrational origin of the coherent oscillations, see Fig. 5. The analogous nonrephasing frequency map can be found in SI.

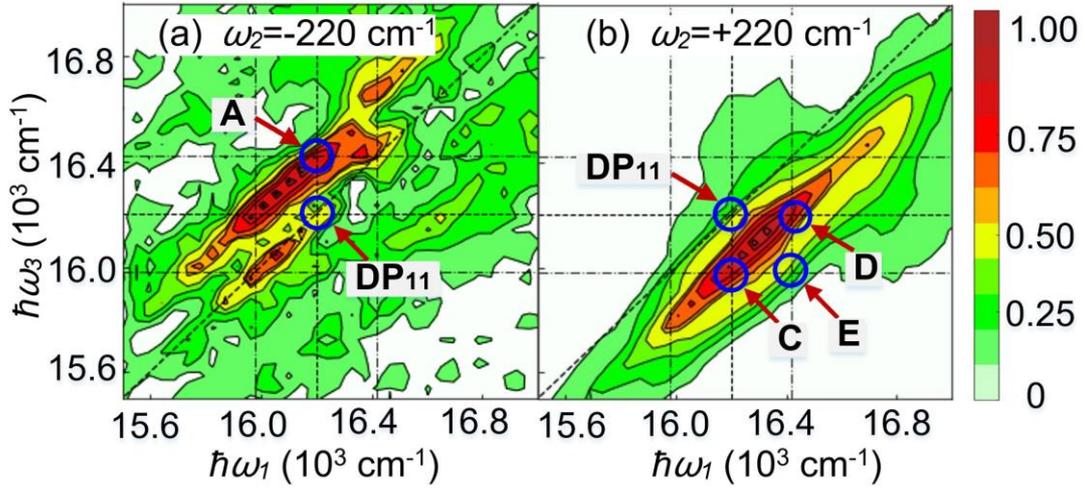

**FIG. 5.** The frequency maps of the ± 220 cm$^{-1}$ frequency components of the rephasing 2D spectra of CdSe QDs. (a). Negative frequency map. The signals at two blue circles A and DP$_{11}$ are explained via the Feynman diagrams of Fig 6. (b). Analogous positive frequency map. The features C, D, E, and DP$_{11}$ are explained via the Feynman diagrams in Fig. 6.

We discuss the frequency maps in terms of Liouville pathways visualized via Feynman diagrams,[21,50–52] some of which are shown in Fig. 6. The upper panels represent rephasing 2D frequency maps. The left panel is for the negative and the right is for the positive beating frequency. The circles with blue, red, and pink outlines in Fig. 6 represent quantum beatings from CdSe QDs with different electronic transition energies. The origin of these differences is the size of the QDs, illustrating how the size distribution gives rise to the elongation of the peaks. The vibrational beatings of the peaks marked with A and DP$_{11}$ at $\omega_2 = -220$ cm$^{-1}$ arise due to the coherence $|e_1, v_1\rangle\langle e_1, v_0|$, shown in Fig. 6 as Feynman diagrams a1 and a2, respectively.

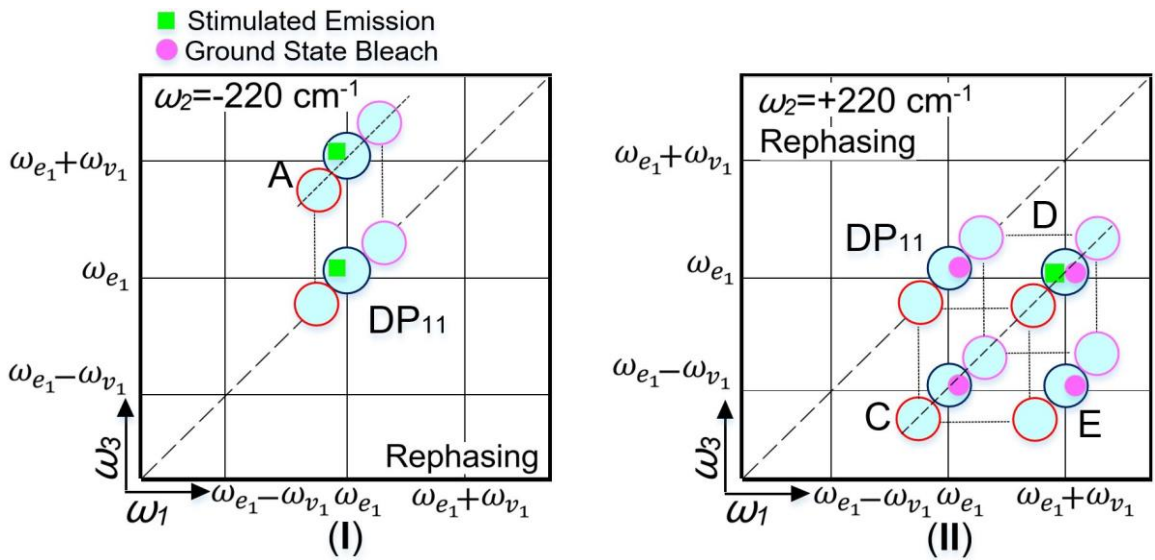

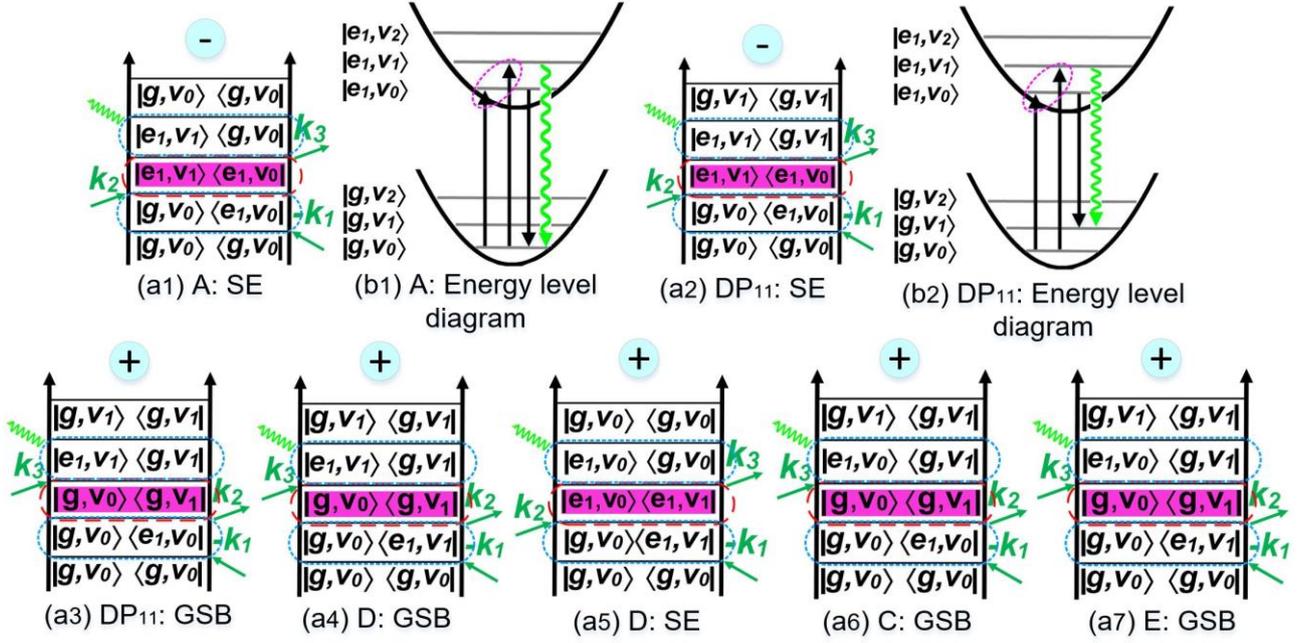

**FIG. 6.** Models of quantum beatings in the rephasing 2D maps ($\omega_2 = \pm 220$ cm$^{-1}$). The green square represents the SE pathway. The pink circle represents the GSB pathway. (I). Model of the 2D map at −220 cm$^{-1}$. (II). Model of the 2D map at +220 cm$^{-1}$. (a). Double-sided Feynman diagrams of the peaks in question. (b). Energy level diagrams of the peaks in question.

At $\omega_2 = +220$ cm$^{-1}$, the quantum beating from the vibrational coherence on the electronic ground state $|g, v_0\rangle\langle g, v_1|$ contribute to the signal at DP$_{11}$-GSB, C-GSB, D-GSB, and E-GSB. These are represented in Fig. 6 (a3, a6, a4, and a7, respectively). To the peak D, the excited state coherence also contributes. The vibrational beatings of the rephasing signal are summarized in Table I.

**TABLE I**. Summary of the vibrational beating of CdSe QDs in the rephasing 2D spectrum (cm$^{-1}$)

| | Vibrational beating (Rephasing) | | | | | |
|---|---|---|---|---|---|---|
| Peaks | A | DP$_{11}$ | D | DP$_{11}$ | C | E |
| $\omega_\tau, \omega_t$ | $\omega_{e_1}, \omega_{e_1}+\omega_{v_1}$ | $\omega_{e_1}, \omega_{e_1}$ | $\omega_{e_1}+\omega_{v_1}, \omega_{e_1}$ | $\omega_{e_1}, \omega_{e_1}$ | $\omega_{e_1}, \omega_{e_1}+\omega_{v_1}$ | $\omega_{e_1}+\omega_{v_1}, \omega_{e_1}-\omega_{v_1}$ |
| $\omega_\tau, \omega_t$ | 16200, 16420 | 16200, 16200 | 16420, 16200 | 16200, 16200 | 16200, 15980 | 16420, 15980 |
| $\omega_T$ | −220 | −220 | +220 | +220 | +220 | +220 |
| pathways | SE | SE | SE & GSB | GSB | GSB | GSB |
| Signal | $|e_1, v_1\rangle\langle e_1, v_0|$ | $|e_1, v_1\rangle\langle e_1, v_0|$ | $|g, v_0\rangle\langle g, v_1|$ & $|g, v_0\rangle\langle g, v_1|$ | $|g, v_0\rangle\langle g, v_1|$ | $|g, v_0\rangle\langle g, v_1|$ | $|g, v_0\rangle\langle g, v_1|$ |
| Beating | Excited-state coherence | | | Ground-state coherence | | |

In order to assess the relative peak amplitudes, we use the four transitions present in the Feynman diagrams. Each transition involves overlap integral of the shifted vibrational states. Using the linear harmonic Franck-Condon model the following analytical expression can be obtained[53]:

$$\langle \chi_{aM} | \chi_{bN} \rangle = \exp[-(\Delta g_{ab})^2/2] \sum_{m=0}^{M} \sum_{n=0}^{N} \frac{(-1)^n (\Delta g_{ab})^{m+n}}{m! \, n!} \times \sqrt{\frac{M! \, N!}{(M-m)! \, (N-n)!}} \, \delta_{(M-m, N-n)}$$

Here $M, N$ and $a, b$ refer to vibrational and electronic states, respectively. $(\Delta g_{ab})^2$ is the Huang-Rhys factor related to the displacement of the nuclear vibrational potentials due to the electronic transition. The only transitions that are important for the analyses involve only $M, N = 0, 1$. Since the

Huang-Rhys factor in QDs is relatively small (~0.1)[54], the strongest transition is the one between the ground vibrational states ($N = 0, M = 0$). The overlap integral of the first vibrationally excited ($N = 1, M = 1$) states is the second largest and identical to the first one with the exception of a factor $(1 - (\Delta g_{ab})^2)$. This means that we can expect in our negative frequency case that peak A should have a larger magnitude than $DP_{11}$, as the pathway contributing to peak A has two (0, 0) transitions, whereas $DP_{11}$ has one (1, 1) and one (0, 0) transition while both pathways have two (0,1) transitions. This indicates that the signal is stronger in the pathway for A and it should be more visible of the two, consistent with the corresponding experimental frequency maps. In the positive frequency case, we can expect that the large peak, which both C and D contribute to should be significantly larger than both peaks $DP_{11}$ and E for the same reason. All positive-frequency pathways have two (0, 1) transitions, where the pathway for C and one of D's pathways each have two (0, 0) transitions in addition to D's second pathway with two (1, 1) transitions. $DP_{11}$ and E have one of each interaction in their pathways and should therefore individually be lower than either C or D. In general, the considerations based on this very simple Franck−Condon model are in good qualitative agreement with the experimental results. For example, the conclusion that the peaks A and C+D should be dominant in the negative and positive frequency cases is in good agreement with the experiment.

Earlier simulations of 2D spectroscopy have shown that the QD size dependence of the LO-phonon frequency would most prominently be visible as a tilt in the phase map of the 2D beating signal.[52] Also the modelled amplitude map shows a tilt, though not as clear. In Fig. 7 we show a 2D map of the amplitude and phase for the $\omega_2 = +220$ cm$^{-1}$ rephasing signal. The corresponding negative frequency and the nonrephasing data are presented in SI. No significant tilt in either the amplitude or phase map can be identified. We conclude that the size dependence of the LO-phonon frequency in the measured CdSe QDs is smaller than the resolution of this study – about 10 cm$^{-1}$.[35]

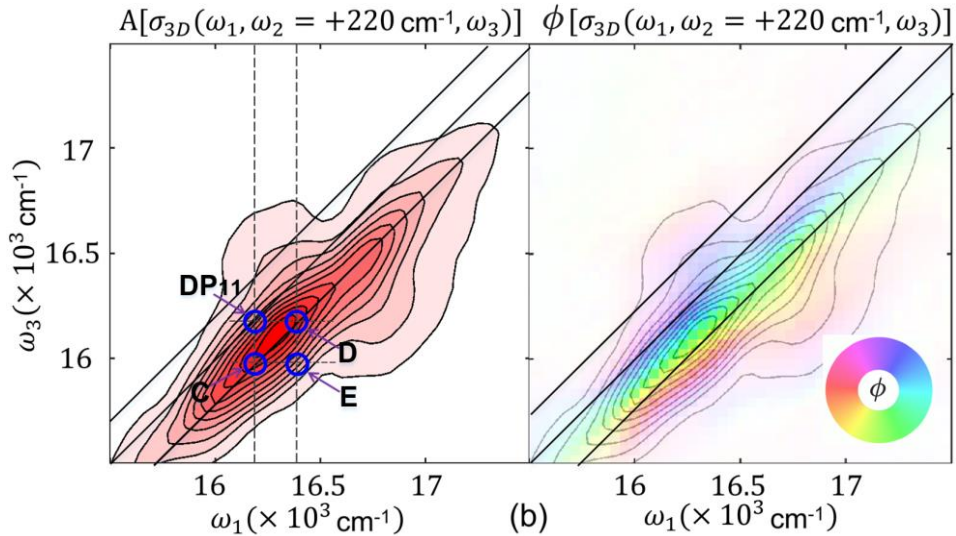

**FIG. 7.** The amplitude and phase frequency maps of the +220 cm$^{−1}$ frequency component of the rephasing 2D spectra of the CdSe QDs. The amplitude plot is displayed on the left and the phase is displayed on the right. Equivalent plot for the -220 cm$^{−1}$ can be found in SI (Fig. S2).

Earlier the combined effect of two vibrational modes, has been analyzed.[28,55] Here too, both LO and LA modes are strongly coupled to the electronic transition. However, at 77 K, the spectrally broad phonon-wing-like LA mode is thermally excited and in this situation would mostly contribute to the line-shape rather than appear as distinct features in the spectra.[56] The same should be valid even for 2D spectroscopy and the frequency maps.

## CONCLUSIONS

We analyzed the beating signals in coherent 2D spectroscopy of CdSe QDs using the double-sided Feynman diagram method. We make use of the fact that the 2D spectra are complex-valued and thereby the Fourier transform provides complementary information for the positive and negative frequencies. We identify the dominant quantum beatings of 210 cm$^{-1}$ in CdSe QDs at 77 K as coherent oscillations of LO phonons. Within uncertainty of the experiment no QD size dependence of the LO phonon frequency was identified. The used methodology is generally suitable for unraveling the complex coherent signals in 2D spectroscopy.

## SUPPLEMENTARY MATERIAL

For specific details, please refer to the supplementary material for a detailed compilation of the relevant experimental equipment, sample preparation, and extension of the experimental results.

## DATA AVAILABILITY

The data that support the findings of this study are available from the corresponding author upon reasonable request.

## ACKNOWLEDGMENTS

We thank Dr. J. O. Tollerud, Dr. E. Bukarte, Dr. J. Luttig, and Dr. P. Kolesnichenko for the help with processing the experimental 3D data. This work was financed by the Swedish Research Council (VR), the Knut and Alice Wallenberg Foundation, and the Swedish Energy Agency. We acknowledge collaboration within NanoLund. This project has received funding from the European Union's Horizon 2020 research and innovation program under the Marie Skłodowska-Curie grant agreement No 945378, and the ADONIS project (CZ.02.1.01/0.0/0.0/16_019/0000789) of the European Regional Development Fund and the Ministry of Education, Youth and Sports of the Czech Republic.

## AUTHOR DECLARATIONS

### Author Contributions

Z. Wang and A. Hedse contributed equally to this work. All authors have read and agree with the final manuscript.

### Corresponding Author

Tõnu Pullerits − Division of Chemical Physics and NanoLund, Lund University, 22100 Lund, Sweden; orcid.org/0000-0003-1428-5564; Email: tonu.pullerits@chemphys.lu.se

### Authors

Zhengjun Wang − Division of Chemical Physics and NanoLund, Lund University, 22100 Lund, Sweden; orcid.org/0000-0002-7599-0382

Albin Hedse − Division of Chemical Physics and NanoLund, Lund University, 22100 Lund, Sweden

Edoardo Amarotti − Division of Chemical Physics and NanoLund, Lund University, 22100 Lund, Sweden.

Nils Lenngren − Division of Chemical Physics and NanoLund, Lund University, 22100 Lund, Sweden; ELI Beamlines, Institute of Physics, Czech Academy of Sciences, 252 41 Dolní Brež any, Czech Republic; orcid.org/0000-0001-7563-9843.

Karel Žídek − Division of Chemical Physics and NanoLund, Lund University, 22100 Lund, Sweden; Regional Center for Special Optics and Optoelectronic Systems (TOPTEC), Institute of Plasma Physics of the Czech Academy of Sciences, 270 00 Prague 8, Czech Republic.

Kaibo Zheng − Division of Chemical Physics and NanoLund, Lund University, 22100 Lund, Sweden; Department of Chemistry, Technical University of Denmark, DK-2800 Kongens Lyngby, Denmark; orcid.org/0000-0002-7236- 1070.

Donatas Zigmantas − Division of Chemical Physics and NanoLund, Lund University, 22100 Lund, Sweden; orcid.org/0000-0003-2007-5256.

Tõnu Pullerits − Division of Chemical Physics and NanoLund, Lund University, 22100 Lund, Sweden; orcid.org/0000-0003-1428-5564; Email: tonu.pullerits@chemphys.lu.se

**Conflict of Interest**

The authors declare no competing financial interests.

# Supplementary Information

# Beating signals in CdSe quantum dots measured by low-temperature 2D spectroscopy

Zhengjun Wang,[1] Albin Hedse,[1] Edoardo Amarotti,[1] Nils Lenngren,[1,2] Karel Žídek,[1,3] Kaibo Zheng,[1,4] Donatas Zigmantas,[1] Tõnu Pullerits[1]

**AFFILIATIONS**

[1] Division of Chemical Physics and NanoLund, Lund University, P.O. Box 124, 22100 Lund, Sweden.
[2] ELI Beamlines, Institute of Physics of the Czech Academy of Sciences, Za Radnicí 835, 252 41 Dolní Břežany, Czech Republic.
[3] Regional Center for Special Optics and Optoelectronic Systems (TOPTEC), Institute of Plasma Physics of the Czech Academy of Sciences, 182 00 Prague 8, Czech Republic.
[4] Department of Chemistry, Technical University of Denmark, DK-2800 Kongens Lyngby, Denmark.

# Contents

# S1. Experiment and Sample Preparation

A detailed description of the set-up used to obtain the 2D spectra is given in Augulis and Zigmantas[1]. Firstly, a 10 kHz Yb: KGW laser generates 1030 nm pulses which enter a non-linear optical parametric amplifier (OPA). The OPA outputs a long pulse which is compressed to 9.2 fs and centered at 600 nm. The pulses are then delayed with a delay stage and a glass wedge. The purpose of this is to vary both the total time and the time-delay between pulses while keeping track of them. The signal beam from the system's response is then detected via heterodyne detection with low-intensity pulses in the phase matching direction. The sample is then probed with pulses with an energy of 1 nJ and a spot diameter of 100 μm. The CCD of the spectrometer then records an interferogram. 2D spectra were obtained using a wide range of coherence times (−141 fs to 360 fs, with steps of 1.5 fs) with population times of up to 20 ps.

The CdSe QDs were synthesized as described in Bullen[2] and Zheng[3]. CdO with oleic acid as a capping agent was heated to 340 °C, then injected into Se with trioctylphosphine as a capping agent into an octadecene solution, and quickly cooled. The sample was then purified and transferred to a hexane solution to grow. Finally, the capping agent was changed to 3-mercaptopropionic acid, and the solvent was changed to ethanol. The quantum dots were measured to have a diameter of 7 nm in a solvent glass. The quantum dots were then re-dispersed in a 1:1 mixture of methanol and ethanol and cooled to 77 K in a cryostat.

# S2. Extension of Experimental Results

The full time-domain scan for a point in the 2D spectrum can be seen in Fig. S1.

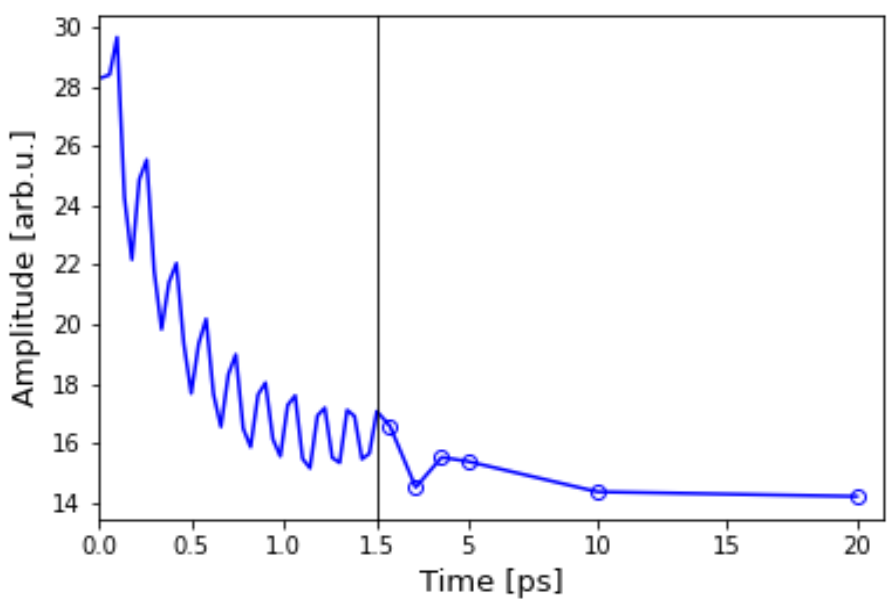

**FIG. S1.** The rephasing signal of the CdSe QDs. ($\omega_1 = \omega_3 = 16200$ cm$^{-1}$) for population times up to 20 ps. Individual points after 1.5 ps are marked with circles.

The quantum beatings in the other 2D slice is shown in Fig. S2 for comparison.

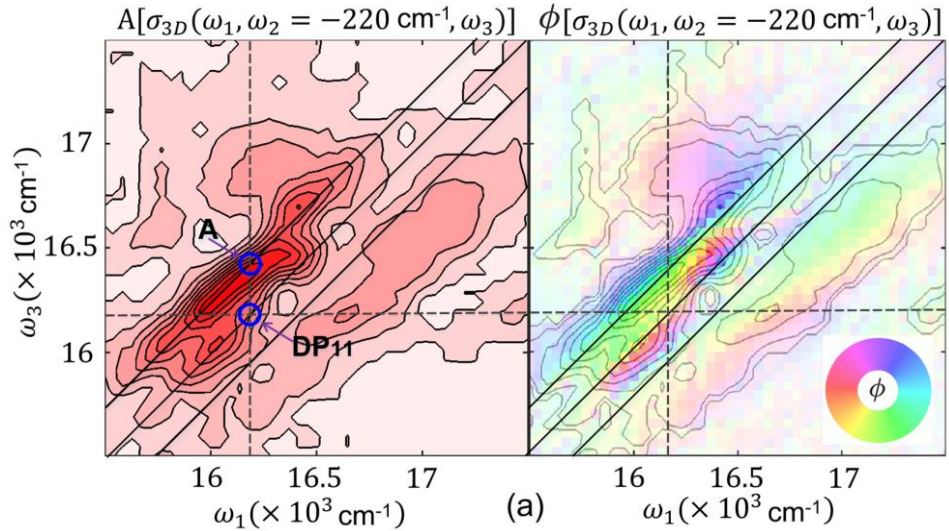

**FIG. S2.** The amplitude part of rephasing 2D spectra of CdSe QDs. ($\omega_2$=-220 cm$^{-1}$)

The photon energies of the pulses used for the 2D analysis range from 15500 cm$^{-1}$ to 18800 cm$^{-1}$. A slice of the 2D absorption spectrum[4] was fitted with peaks representing the QD states, this is shown in Fig. S3.

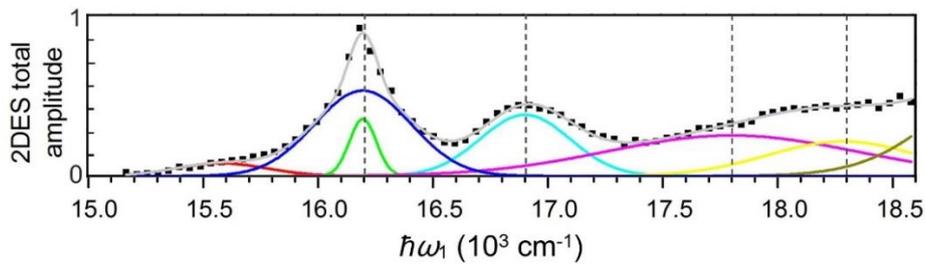

**FIG. S3**. A slice of the total 2D absorption spectrum is fitted to locate the states of the QDs at $\omega_3$=16200 cm$^{-1}$ at a population time of 10 ps. The black squares represent the measured data with the colored peaks corresponding to specific states, the sum of these peaks is marked with a gray line. The average energy of each state is marked with a vertical dashed line.

The vibrational beatings in the nonrephasing spectrum at ±220 cm$^{-1}$ were also analyzed, shown in Fig. S4. The nonrephasing component is generally weak and the signal to noise ratio is low. Consequently, the agreement of the peaks at −220 cm$^{-1}$ and +220 cm$^{-1}$ with the theoretical prediction is poor.

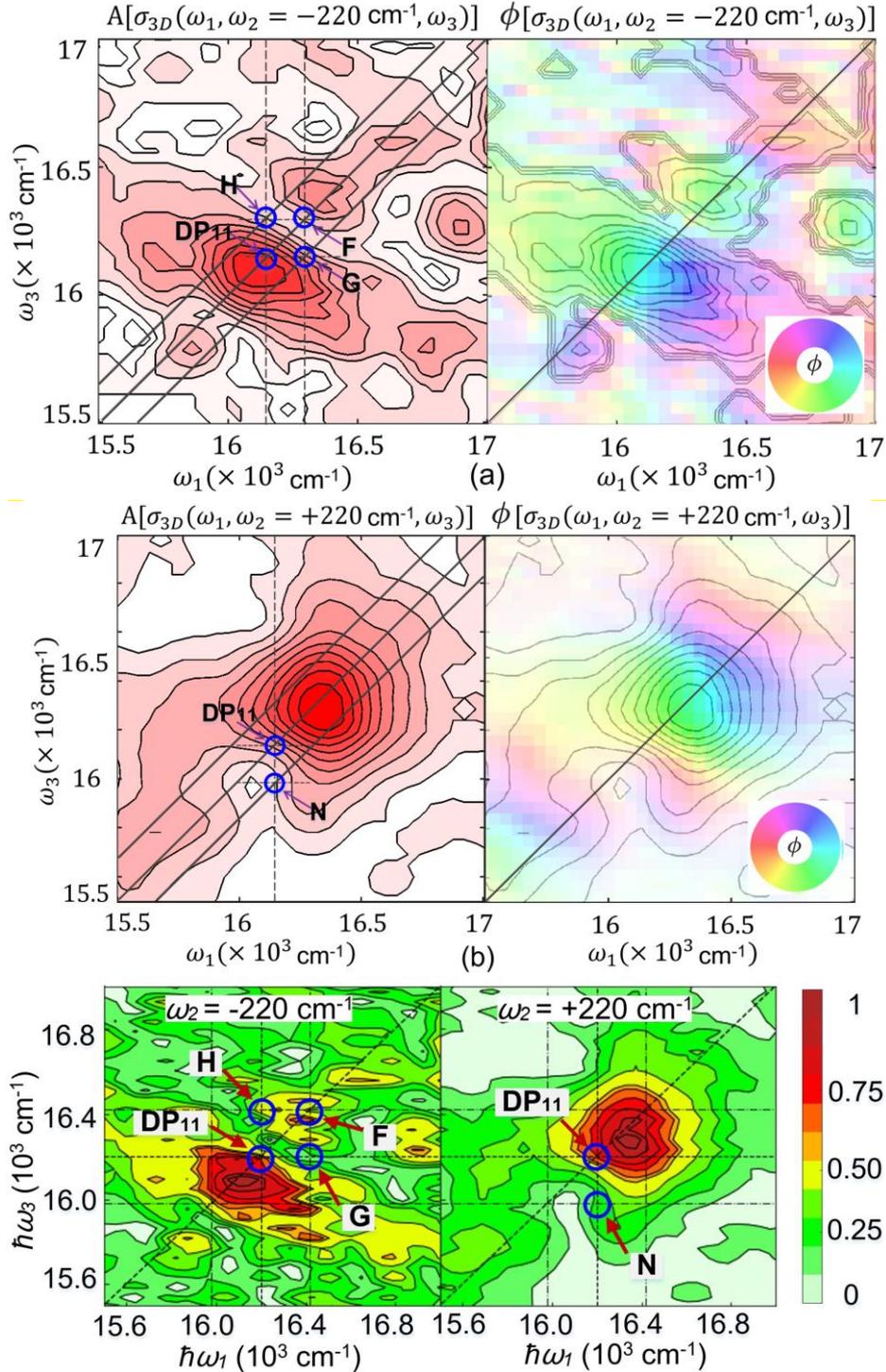

**FIG. S4.** The amplitude part of non-rephasing 2D spectra of CdSe QDs. ($\omega_2=\pm220$ cm$^{-1}$). (a). Four vibrational beating peaks DP$_{11}$, F, G, and H of the 2D map at $-220$ cm$^{-1}$. (b). Two vibrational beating peaks DP$_{11}$ and N of the 2D map at $+220$ cm$^{-1}$.

The theoretical model used describes the vibrational beating profile for the nonrephasing component in Fig. S5. The predicted beating model is summarized in Table S1 for the 2D map at $-220$ cm$^{-1}$ in Fig. S5(a1), the contribution of vibrational beating comes from the ground state coherence $|g, v_1\rangle\langle g, v_0|$ in DP$_{11}$-GSB. The corresponding energy level diagram is shown in Fig. S5(b1). The vibrational beatings in the N peak mainly come from the excited state coherence of $|e_1, v_1\rangle\langle e_1, v_0|$ in Fig. S5(a2) and (a3), with the relative energy level diagrams in Fig. S5(b2) and (b3). The related model for these quantum beatings is shown in Fig. S5(I).

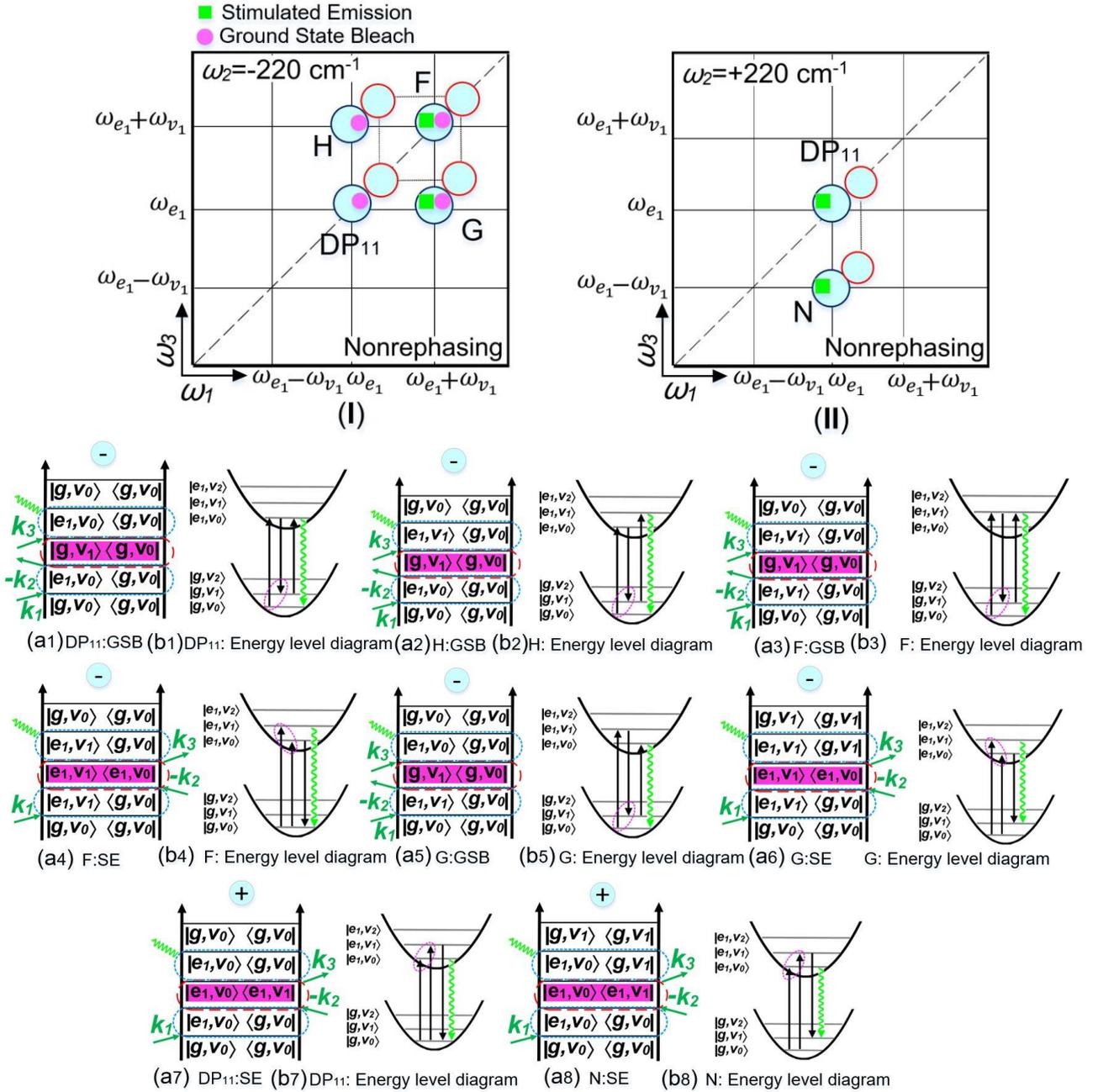

**FIG. S5.** The models of quantum beatings in non-rephasing 2D maps ($\omega_2=\pm220$ cm$^{-1}$). The green square represents the SE pathway. The pink diamond stands for the GSB pathway. (I). The model of 2D map at −220 cm$^{-1}$. (II). The model of 2D map at +220 cm$^{-1}$. (a). Double-sided Feynman diagrams of the peaks in question. (b). Energy-level diagrams of the peaks in question.

For the 2D maps at +220 cm$^{-1}$ in Fig. S5(II), the vibrational beating of DP$_{11}$ comes from the $|e_1, v_0\rangle\langle e_1, v_1|$ coherence in the DP$_{11}$-SE pathway with corresponding energy level diagram in Fig. S5(b4). Similarly, the vibrational beating of H comes from the coherence $|e_1, v_0\rangle\langle e_1, v_1|$, shown in Fig. S5(a5) with its corresponding energy level diagram in Fig. S5(b5).

**TABLE S1.** Summary of the theoretical vibrational beating of CdSe QDs in the non-rephasing 2D spectrum

| | Vibrational beating (Non-rephasing) | | | | |
|---|---|---|---|---|---|
| Peaks | DP$_{11}$ | H | F | DP$_{11}$ | G |
| $\omega_\tau, \omega_t$ | $\omega_{e_1}, \omega_{e_1}$ | $\omega_{e_1}, \omega_{e_1}+\omega_{v_1}$ | $\omega_{e_1}+\omega_{v_1}, \omega_{e_1}+\omega_{v_1}$ | $\omega_{e_1}, \omega_{e_1}$ | $\omega_{e_1}+\omega_{v_1}, \omega_{e_1}$ |
| $\omega_\tau, \omega_t$ (cm$^{-1}$) | 16200, 16200 | 16200, 16420 | 16420, 16420 | 16200, 16200 | 16420, 16200 |
| $\omega_T$ (cm$^{-1}$) | $-220$ | $-220$ | $-220$ | $+220$ | $-220$ |
| Pathways | GSB | GSB | GSB & SE | SE | SE & GSB |
| Signal | $|g,v_1\rangle\langle g,v_0|$ | $|g,v_1\rangle\langle g,v_0|$ | $|g,v_1\rangle\langle g,v_0|$ & $|e_1,v_1\rangle\langle e_1,v_0|$ | $|e_1,v_0\rangle\langle e_1,v_1|$ | $|e_1,v_1\rangle\langle e_1,v_0|$ & $|g,v_1\rangle\langle g,v_0|$ |
| Beating | Ground-state coherence | Excited-state coherence | | Excited-state coherence | |

The untreated version of Fig. 4 from the main text is shown in Fig. S6. Before the Fourier transform was performed, the exponential decay and a constant background were subtracted from the time domain signal in order to remove the DC component and clarify the peak location, as the center was obscured due to its Fano shape. The time domain signal was then also zero-padded and windowed to smoothen the frequency curve.

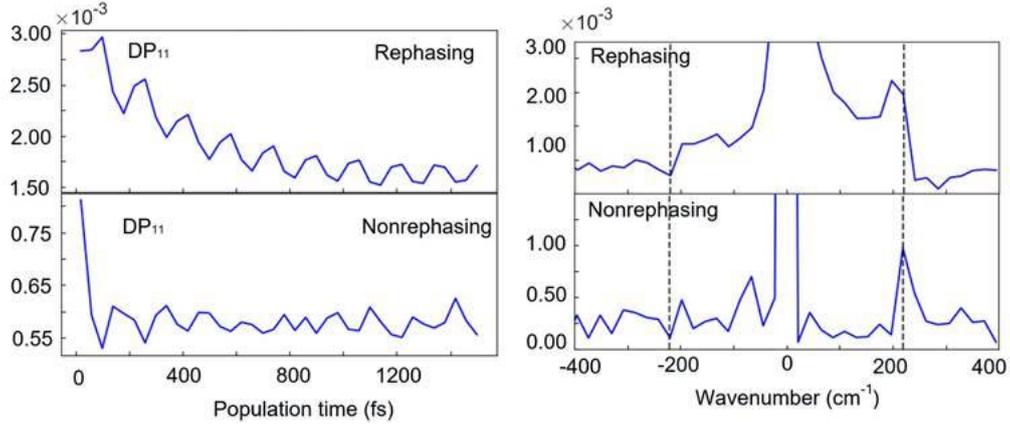

**FIG. S6.** Left: The population time dependence of the rephasing (upper) and nonrephasing (lower) signal at the position of DP$_{11}$. Right: the Fourier transform of the time domain signal, with the data points at ±220 cm$^{-1}$ marked by a vertical dashed line.

A 1D-slice comparison of phase and amplitude is shown in Fig. S7.

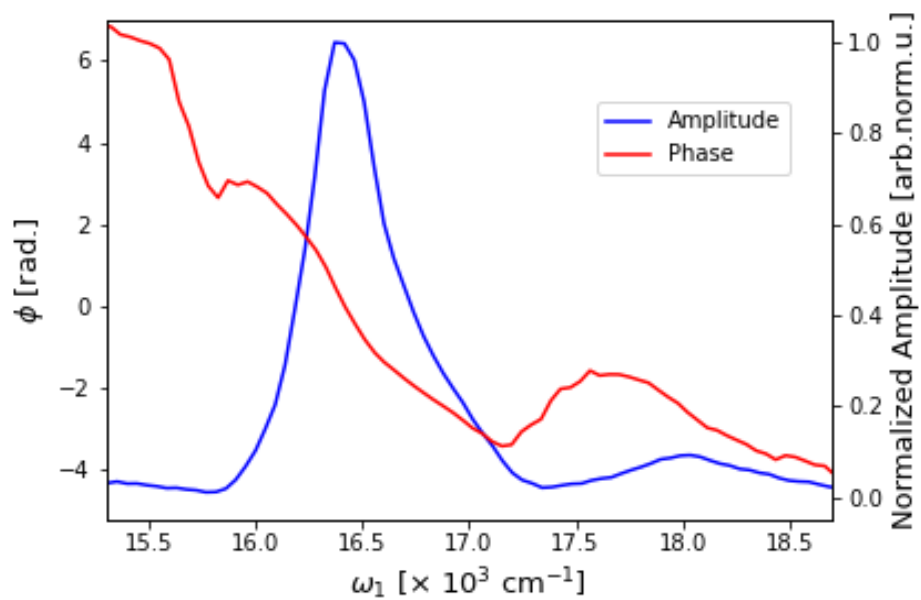

**FIG. S7.** The amplitude and phase parts from the rephasing 2D spectra of CdSe QDs. ($\omega_3$=16200 cm$^{-1}$)